\documentclass[12pt,preprint]{aastex}

\shorttitle{G- and P-modes in  HD 163899}
\shortauthors{Saio et al.}

\begin{document}


\title{{\it MOST}\footnotemark[1]~ DETECTS G- AND P-MODES 
IN THE B SUPERGIANT HD 163899 (B2Ib/II)}

\footnotetext[1]{Based on data from the {\it MOST} satellite, a Canadian Space Agency mission, jointly operated by Dynacon Inc., the University of Toronto Institute of
Aerospace Studies and the University of British Columbia with the assistance of the University of Vienna.} 

\author{H. Saio\altaffilmark{2}, 
R. Kuschnig\altaffilmark{3},  
A. Gautschy\altaffilmark{4}, C. Cameron\altaffilmark{3}, G.A.H. Walker\altaffilmark{5}, J.M. Matthews\altaffilmark{3}, 
D.B. Guenther\altaffilmark{6}, A.F.J. Moffat\altaffilmark{7}, 
S.M. Rucinski\altaffilmark{8}, D. Sasselov\altaffilmark{9}, 
W.W. Weiss\altaffilmark{10}
} 

\altaffiltext{2}{Astronomical Institute, Graduate School of Science, Tohoku University, Sendai, 980-8578, Japan;
saio@astr.tohuku.ac.jp}

\altaffiltext{3}{Dept. Physics and Astronomy, UBC, 
6224 Agricultural Road, Vancouver, BC V6T 1Z1, Canada;  kuschnig@astro.phys.ubc.ca, matthews@phas.ubc.ca, ccameron@phas.ubc.ca}

\altaffiltext{4}{Wetterchr\"uzstr. 8c, CH-4410 Liestal, Switzerland;
alfred@gautschy.ch}

\altaffiltext{5}{1234 Hewlett Place, Victoria, BC V8S 4P7, Canada;
gordonwa@uvic.ca}

\altaffiltext{6}{Department of Astronomy and Physics, St. Mary's University
Halifax, NS B3H 3C3, Canada;
guenther@ap.stmarys.ca}

\altaffiltext{7}{D\'ept. de physique, Univ. de Montr\'eal 
C.P.\ 6128, Succ.\ Centre-Ville, Montr\'eal, QC H3C 3J7, Canada;
and Obs. du mont M\'egantic;
moffat@astro.umontreal.ca}

\altaffiltext{8}{Dept. Astronomy \& Astrophysics, David Dunlap Obs., Univ. Toronto 
P.O.~Box 360, Richmond Hill, ON L4C 4Y6, Canada;
rucinski@astro.utoronto.ca}

\altaffiltext{9}{Harvard-Smithsonian Center for Astrophysics, 
60 Garden Street, Cambridge, MA 02138, USA;
sasselov@cfa.harvard.edu}

\altaffiltext{10}{Institut f\"ur Astronomie, Universit\"at Wien 
T\"urkenschanzstrasse 17, A--1180 Wien, Austria;
weiss@astro.univie.ac.at}

\begin{abstract}
The {\it Microvariability and Oscillations of Stars (MOST)} satellite
observed the B supergiant HD 163899 (B2 Ib/II) for 37 days as a guide
star and detected 48 frequencies  $\la$ 2.8 c~d$^{-1}$
with amplitudes of a few milli-magnitudes (mmag) and less.  The frequency range
embraces g- and p-mode pulsations.  It was generally thought that no
g-modes are excited in  less luminous B supergiants because strong 
radiative damping is expected in the core.  
Our theoretical models, however, show that
such g-modes are excited in massive post-main-sequence stars, in
accordance with these observations.  
The nonradial pulsations excited in models between
$20M_\odot$ at $\log T_{\rm eff} \approx 4.41$ and 
$15M_\odot$ at $\log T_{\rm eff} \approx 4.36$ are roughly consistent with
the observed frequency range.
Excitation by the Fe-bump in opacity
is possible because g-modes can be partially reflected at a convective
zone associated with the hydrogen-burning shell, which significantly
reduces radiative damping in the core.  The {\it MOST} light curve of
HD 163899 shows that such a reflection of g-modes actually occurs,
and reveals the existence of a previously unrecognized type of
variable, slowly pulsating B supergiants (SPBsg) distinct from $\alpha$ Cyg 
variables.  Such g-modes have great potential for asteroseismology.
\end{abstract}

\keywords{stars: individual (HD 163899) -- stars: early-type 
  -- stars: oscillations -- supergiants}

\section{INTRODUCTION}
The B star region in the HR diagram contains a wide range
of variables.  The most luminous  (luminosity classes of 0-Ia;
$\ga 10^5 L_\odot$) are known as $\alpha$ Cyg variables and
show semi-regular light and radial-velocity variations on
timescales of 1 -- 6 weeks \citep[e.g.,][]{van98}.  Those
variations are thought to be a combination of pulsation and rotational
modulation of winds \citep[e.g.,][1997]{van91,Kau96}.  Pulsations in such 
very luminous stars can
be excited mainly by the Fe-bump of opacity \citep{Rog92} at a
temperature of about $2\times10^5$K.  The excitation mechanism is,
however, more affected by the strange-mode character in more luminous
stars \citep{Kir93,Gla96,Dor00}. The target star discussed in this
paper, HD163899 (B2Ib/II), is a supergiant that is less
luminous and less affected by winds than the $\alpha$ Cyg variables.

There are also many less luminous B-type variable stars whose 
light and radial-velocity variations are caused by radial and 
nonradial pulsations. These pulsations are excited by
the classical $\kappa$-mechanism at the Fe-bump of opacity. 
The type of mode excited depends mainly on the effective temperature
of the stars; radial and nonradial p-mode pulsations are excited in 
hotter stars and nonradial g-mode pulsations are excited in relatively 
cooler stars \citep[see e.g.,][]{Pam99}.
Most of the stars are main-sequence stars.
Hotter p-mode pulsators ($\beta$ Cephei stars) are more luminous or
more massive ($\ga 8M_\odot$) than the relatively cooler (B3 -- B8)
g-mode pulsators, slowly pulsating B (SPB) stars \citep{Wae91}. 
Mainly based on line-profile variations, g-modes are known to be
excited also in early-type Be stars \citep[e.g.,][]{Riv03}.

The $\beta$ Cephei stars have been known for more than a century, but
the excitation mechanism was identified only in the early 1990s 
\citep{Kir92,Mos92} after the emergence of OPAL opacity tables
\citep{Rog92}. The observational properties of the $\beta$ Cephei stars 
are thoroughly discussed by \citet{Sta05}. 

The SPB stars are nearly perfectly confined to the main-sequence band
\citep[e.g.,][]{deC04}, in accordance with the theoretical prediction
that the radiative core in a post-main-sequence star strongly damps
g-modes where they have very short wavelengths  \citep{Gau93,Dzi93}.
In this paper, however, we show that reflection of pulsation modes
at the convective zone associated with the hydrogen-burning shell
quenches the radiative damping in the core, and hence the Fe-bump can
excite some g-modes in a post-main-sequence massive star with a radiative
core.

The {\it Microvariability \& Oscillations of STars (MOST)} photometric
satellite was launched on June 2003 and is fully described by
\citet{MOST}.  The first scientific results were published by
\citet{Mat04}.

{\it MOST} observations have significantly clarified the properties of
pulsating B stars.  {\it MOST} detected clear light variations in the
rapidly rotating O9Ve star $\zeta$ Oph \citep{Wal05a}, which revealed
the presence of $\beta$ Cephei-type pulsations in this line-profile
variable.  {\it MOST} found that the mono-periodic $\beta$ Cephei star
$\delta$ Ceti is actually multiperiodic with three additional
frequencies \citep{Aer06a}.  {\it MOST} discovered a new SPB star, HD
163830 (B5II/III), near the edge of the main-sequence band, detecting
20 g-mode frequencies \citep{Aer06b}.  {\it MOST} detected g-modes and
possibly r-modes in the rapidly rotating Be star HD 163868
\citep{Wal05b}. The latter observations showed, for the first time, that many
g-modes can be excited in a rapidly rotating Be star, and provided a
natural interpretation of $\lambda$ Eri-type variables as SPBe stars.
This finding suggests that nonradial pulsations might play a crucial
role in Be phenomena.

In this paper, we report another important discovery by the {\it MOST} 
satellite~--~g-modes in the B supergiant HD 163899.

\section{THE {\it MOST} OBSERVATIONS}
HD 163899 ($V$=8.3, B2 Ib/II) was observed as one of 20
guide stars for the photometry of WR 103 (HD 164270) by the {\it MOST}
satellite.  The observing conditions and the data reduction method were
the same as for the SPBe star HD 163868 described in \citet{Wal05b}.
Observations were made from 2005 June 14 to July 21 for a total of
36.6 days with $>137,000$ values recorded for HD 163899.  The data are
binned in 2 minute intervals. 
Since the WR 103 field is outside the
{\it MOST} Continuous Viewing Zone, the duty cycle was limited to
about 50\% of the 101 min orbit.

\begin{figure}
\epsscale{0.8}
\plotone{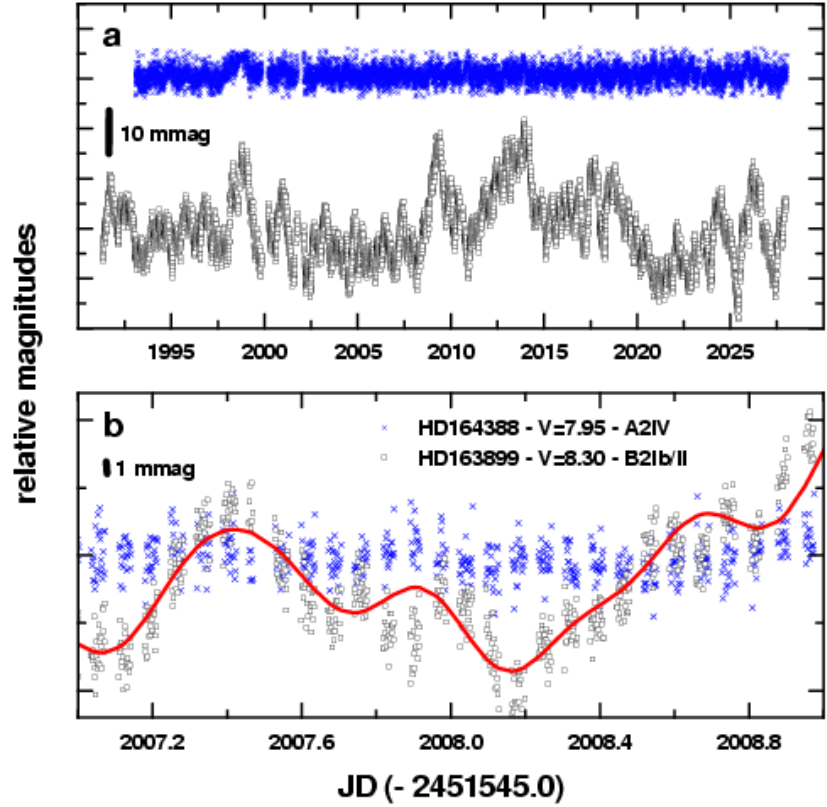}
\caption{{\bf a.} The 2005 {\it MOST} 37 day light curve of the guide star 
HD 163899 and the  simultaneously observed guide star HD 164388 whose
light curve is shifted by 30 mmag for better visibility; {\bf b.} 
a  2 day portion of the HD 163899 light curve with the data binned every 2 min. 
The 2$\sigma$ errors are  $\sim 1.3$ mmag.
Because {\it MOST} was operating outside the continuous viewing zone the duty 
cycle was $\sim$50\% accounting for the gaps in the data. 
The solid line in the bottom panel is the fit of the  48 frequencies detected 
to the data. 
A short term variability of $\sim 0.05$d seen in both stars arose from 
the orbital phase of the
{\it MOST} satellite, and was not included in our frequency search.
}
\label{fig_lc}
\end{figure}

Figure~\ref{fig_lc} displays the full light curve of HD 163899 and for another
guide star, HD 164388 (A2 IV, $V$=8.0), which appears to be constant
(standard deviation $\sim 1.6$ mmag). 
An expanded two day portion of the {\it MOST} light curve is included 
to better display the higher frequency variations. 
The light curve shows 
many periodicities ranging from about a day to a few
days, indicating that many g-mode pulsations are excited 
in HD163899. Complete light curves for both stars can be downloaded 
from the {\it MOST} Public Archive at www.astro.ubc.ca/MOST.

\subsection{The Frequency analysis}

The frequency analysis was carried out using CAPER, a collection of
Fortran driver routines which use a Discrete Fourier Transform (DFT)
as a frequency and amplitude identification tool and non-linear least
squares (NLLS) software \citep[Numerical Recipes;][]{Pres86} 
to refine the identified parameters. The fitting function is a series of sinusoids 
\citep[see][]{Wal05b}.

Periodicities were identified and refined as follows:

a) A frequency/amplitude was identified from the DFT. The amplitude and
phase were refined via NLLS. The fit function was subtracted from the data
and a new set of parameters was identified from the residuals. The new
amplitude and phase was refitted along with all other previously
identified amplitudes and phases. The new function was then subtracted from
the original time series and new parameters identified. This process
is repeated until the next identified frequency has a S/N $\sim 3.6$
($\ga 3$ sigma detection) (Kuschnig et al. 1997).

Noise is defined as the mean of the amplitude spectrum $\pm 3$
c~d$^{-1}$ centered on the identified frequency. The amplitudes are 3 sigma
clipped until the mean converges or until the noise of the entire
frequency range is reached.  
(The frequency range between 0 and 10 c~d$^{-1}$ has a noise of 0.22 mmag.) 

b) Once the 48 frequencies were identified, the three lowest frequencies,
along with their amplitudes and phases were fitted simultaneously by NLLS.

c) The parameters from b) were fixed and all other frequencies,
amplitudes, and phases were simultaneously fitted to the original
light curve. 

\begin{figure}
\epsscale{1.0}
\plotone{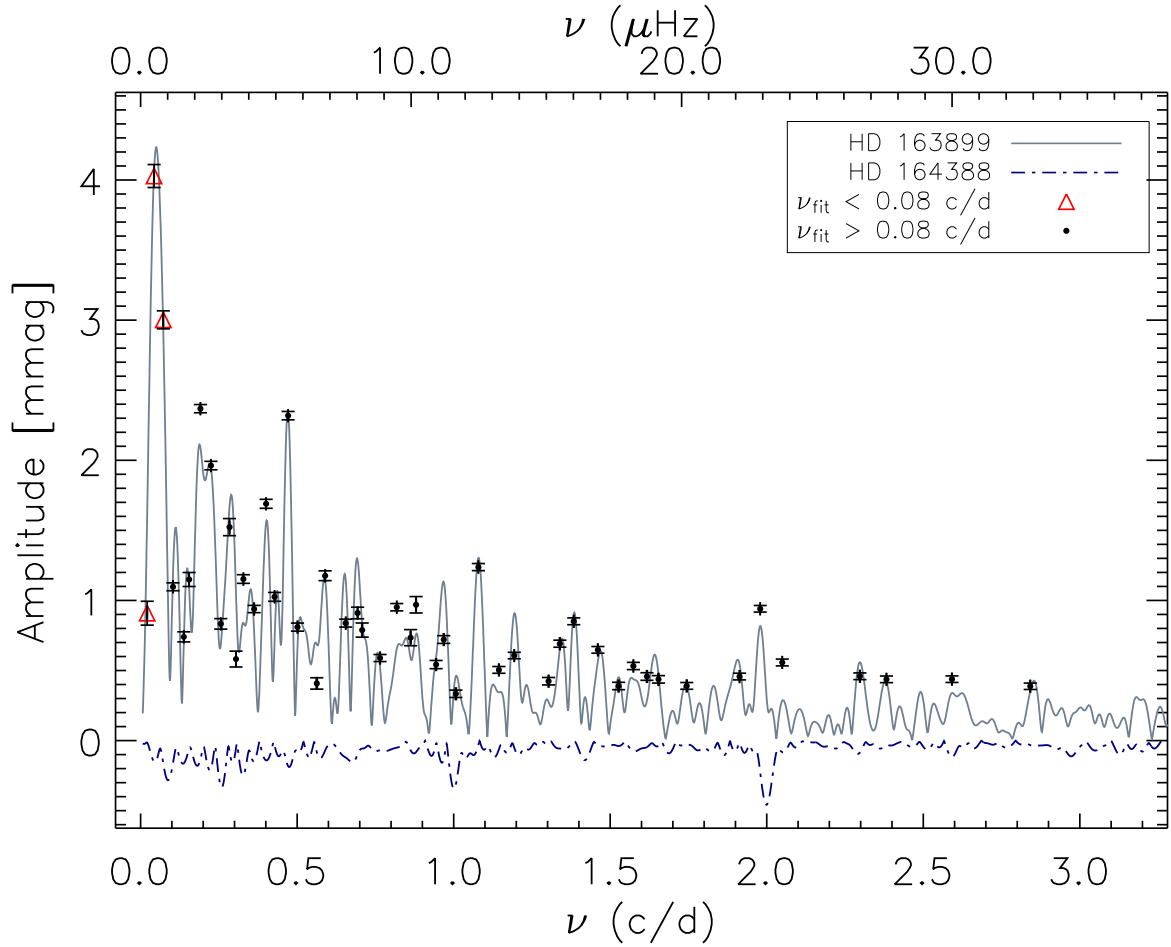}
\caption{
Amplitude spectra from discrete Fourier transforms for the light curves
of HD 163899 (solid line) and the comparison star HD 164388 (dashed line), 
where the sign of the amplitude of the comparison
star is inverted for better visibility.
Dots and triangles with error bars indicate the frequency/amplitude obtained
for HD 163899.
Peaks at 1 and 2 c~d$^{-1}$ conspicuous in the comparison star are 
artifacts typical of the {\it MOST} photometry.
}
\label{amp_sp}
\end{figure}

We separate the fitting of the parameters in b) and c) to
ensure convergence of the fit. This procedure is similar to low pass filtering of the data (i.e. removal of as
much of the long term trend as possible before proceeding). The
parameters from (b) are necessary to improve the residuals. The
results are presented in Table 1. Figure~\ref{amp_sp} show the DFT and the DFT of the comparison
star and the fit parameters with bootstrapped errors (described in the
following paragraph). The lowest frequency f$_{3}$ in Table 1 has 
a period of 47 days which is longer than the data and likely to be less accurate.

\begin{figure}
\epsscale{0.8}
\plotone{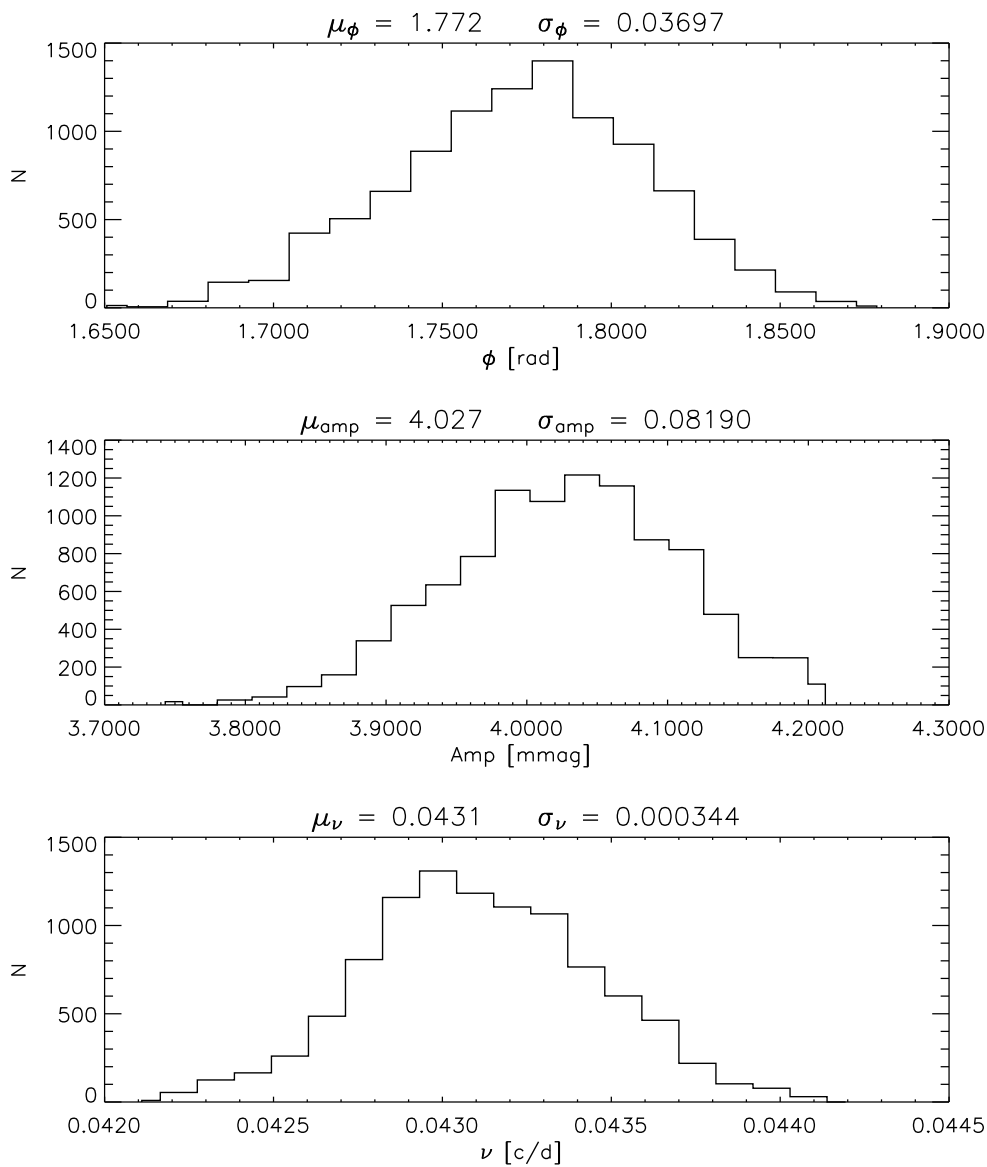}
\caption{
The bootstrap distributions of parameters for the frequency f$_1$. 
They represent the number of the cases at each parameter bin 
resulted from 10,000 light curves produced by inserting random gaps 
into the original. At the top of each diagram the mean value
$\mu$ and the standard deviation $\sigma$ of the distribution are
indicated. Each diagram from the bottom to the top shows the distribution
for the frequency $\nu$ in c~d$^{-1}$, the amplitude in mmag, and the phase $\phi$
in radian.}
\label{dist}
\end{figure}

To assess the precision of our parameter-fitting method in steps (b) and (c)
we use a bootstrap technique \citep[see e.g.][]{Wall03}.
The procedures (b) and (c) are (separately) repeated 10,000
times with a new light curve that is derived from points randomly selected
from the original light curve. The new light curve has the same number of
data points as the original with the possibility that some points may be
repeated. This essentially introduces random gaps in the data. 
Figure~\ref{dist} shows a typical distribution for a parameter set. The sigma of the
distribution is used to estimate the 1-sigma error bars of each parameter.

It is reasonable to assume that these multiple periodicities are
nonradial pulsations simultaneously excited in the star. The analysis
has revealed, in addition to the dense low-frequency spectrum, the
presence of relatively high frequencies up to 2.8 c~d$^{-1}$ which
correspond to p-mode pulsations (i.e.  both p- and g-modes are excited
simultaneously in HD 163899).  To the authors' knowledge, it is the
first clear detection of g-modes in a less luminous B
supergiant, although \citet{Wae98}'s list of the Hipparcos
$\alpha$ Cyg-type supergiants contains some less luminous supergiants
with periods of a few days indicating the possible presence of
g-modes.

\section{MODELING THE OSCILLATIONS}
To examine the stability of post-main-sequence models corresponding
to B supergiants and bright giants, we have computed evolution
models for a mass range of $7 \le M/M_\odot \le 20$ with an initial
chemical composition of $(X,Z)=(0.7,0.02)$.  We used a Henyey-type
stellar evolution code with OPAL opacities \citep{Igl96}.  The
stability of nonradial pulsations has been examined for selected
models by using a finite difference method \citep{Sai80}.  The
stability of a selected model has also been analyzed with a Riccati
shooting code \citep{Gau96}.  We have confirmed that the results of
the two methods agree well.

\begin{figure}
\epsscale{1.0}
\plotone{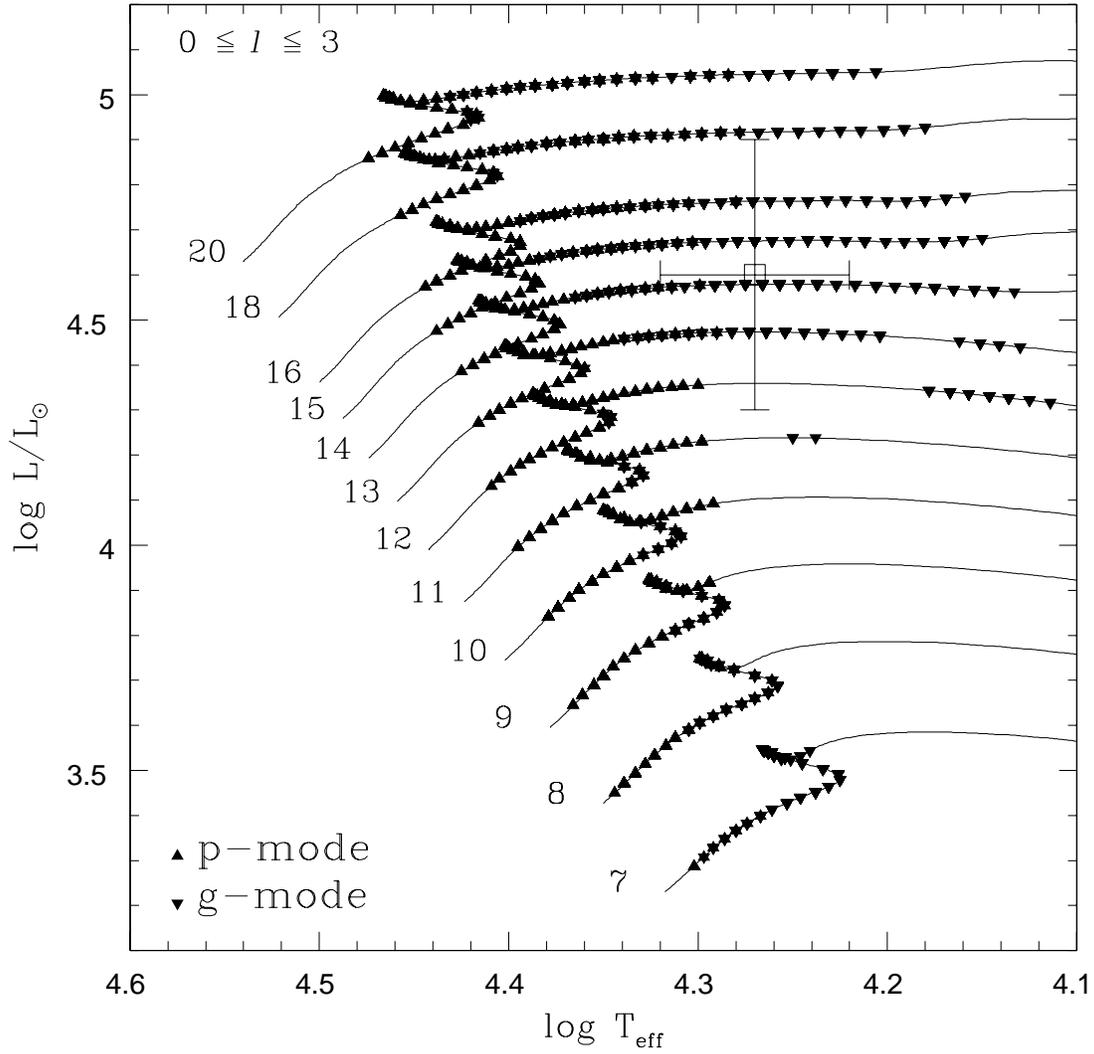}
\caption{Massive star models which excite p-modes (triangles) and
g-modes (inverted triangles) are shown along the evolutionary track
for each mass. The models which excite both g- and p-modes appear as asterisks.
A large square with error bars indicates the position 
of HD 163899 estimated from the spectral type B2 Ib/II.
} 
\label{hrd}
\end{figure}

\begin{figure}
\epsscale{1.0}
\plotone{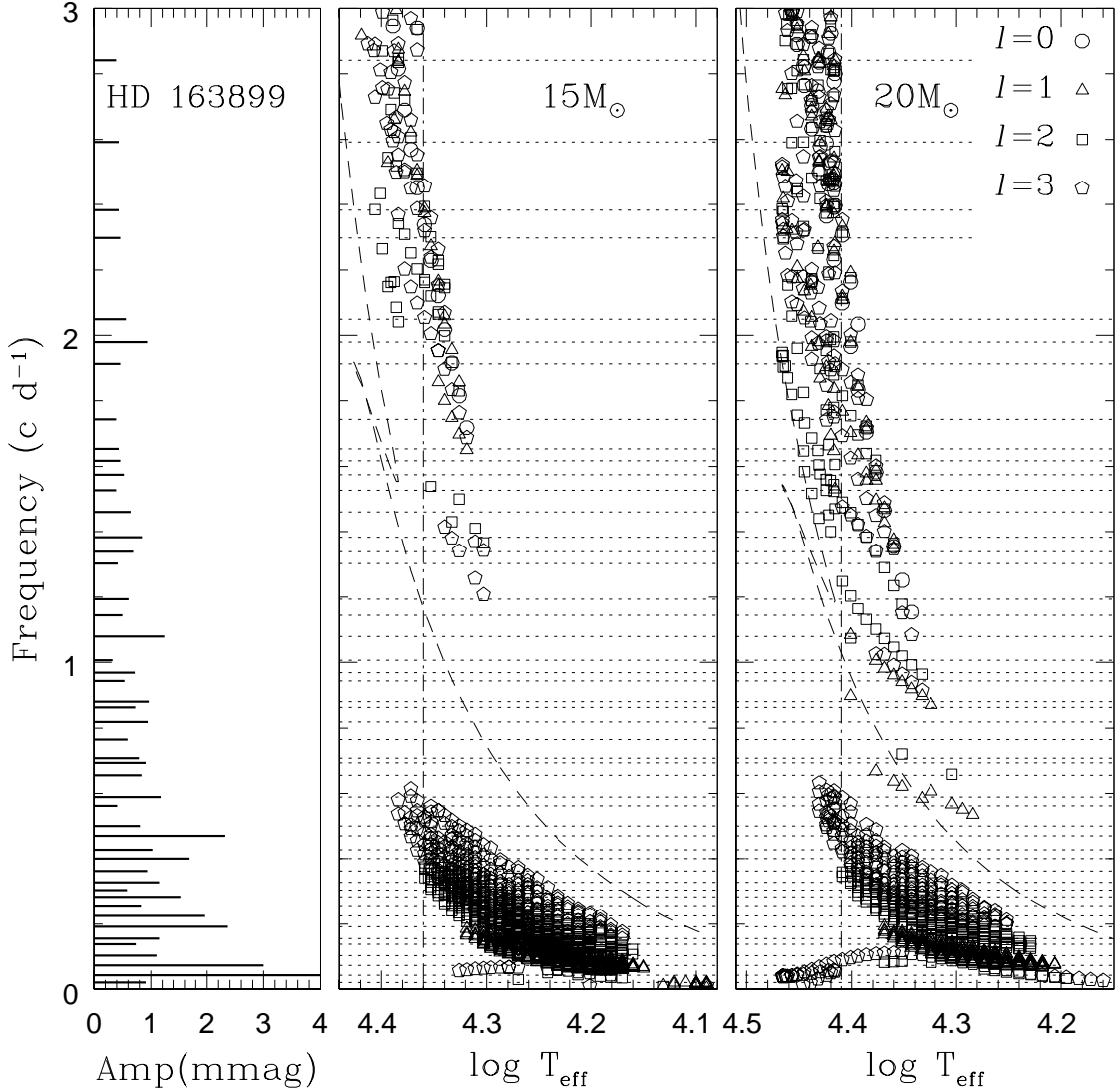}
\caption{
The frequencies of excited pulsations of $0 \le l \le 3$ in evolutionary
models of $15M_\odot$ and $20M_\odot$ against the effective temperature.
The left panel shows frequencies and amplitudes of pulsations obtained
by {\it MOST} for HD 163899.
The horizontal dotted lines in the middle and the right panels
correspond to the {\it MOST} frequencies. 
Long dashed lines in the middle and the right panels indicate
the relation of $\sigma = \sqrt{4\pi G\overline\rho}$, which
is used to discriminate between p- and g-modes in Fig.~\ref{hrd},
where $\sigma$ is the angular frequency of pulsation, $G$ is the 
gravitational constant, and $\overline\rho$ is the mean density of the star.
A vertical dash-dotted line in each of the middle and the right
panels indicates a value of $\log T_{\rm eff}$
around which the frequencies excited in the models are 
roughly consistent with the observed frequencies.
}
\label{te_nu}
\end{figure}

Figure~\ref{hrd} shows positions of models in the HR diagram which
excite p-modes (triangles) and g-modes (inverted triangles).  (The
models which excite both g- and p-modes appear as asterisks.)  If the
angular frequency of pulsation is larger (smaller) than $\sqrt{4\pi
G\overline\rho}$, the mode is classified as a p-mode (g-mode), where
$G$ and $\overline\rho$ are the gravitational constant and the mean
density of the model (see also Fig.\ref{te_nu}).  The p-mode ($\beta$
Cephei-type) instability range and the usual SPB g-mode instability
range for main-sequence stars of $M\la 9M_\odot$ agree with those
obtained by \citet{Pam99}.  We also note that the red edge of the
radial-mode instability range of $20M_\odot$ models is located at
$\log T_{\rm eff} = 4.34$ in perfect agreement with the result of
\citet{Kir93} who examined the stability of radial modes for 
very massive stars.  

In addition to the known instability regions, Figure~\ref{hrd} shows
a wide g-mode instability region for the post-main-sequence massive
($M\ga 12M_\odot$) models.  The instability region has not been
recognized before.  In fact, the stability of g-modes in the
post-main-sequence massive stars had not been examined since the
emergence of OPAL opacities.  [\citet{Gla96} examined the stability of
p-modes but not of higher-order g-modes.] In the next section, we
focus on the the properties of unstable high-order g-modes.  

In the hotter part of the instability region p-modes are also excited
simultaneously.  The position of HD 163899 shown in Figure~\ref{hrd},
which is roughly estimated from its spectral type B2 Ib/II, indicates
that HD 163899 lies in the newly found instability region.  In other
words, HD 163899 belongs to a previously unrecognized type of
pulsating stars, which we call ``slowly-pulsating B-supergiants'' (or
SPBsg).  Strictly speaking, HD 163899, showing both g- and p-modes, is
a hybrid of SPBsg and $\beta$ Cephei-type.

\begin{figure}
\epsscale{1.0}
\plotone{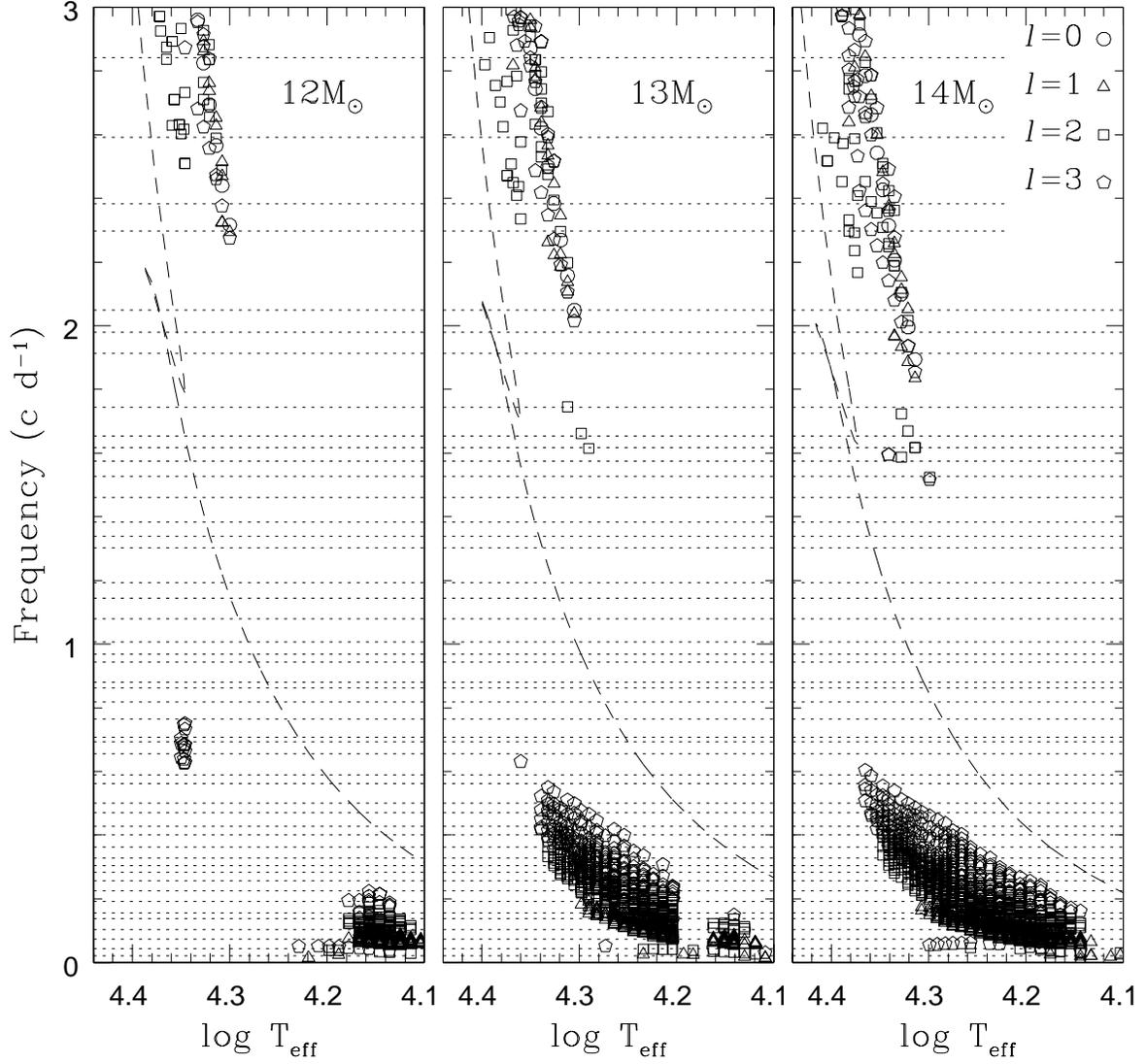}
\caption{
The frequencies of excited pulsations of $0 \le l \le 3$ in evolutionary
models of $12M_\odot$, $13M_\odot$  and $14M_\odot$ against 
the effective temperature.
The dashed lines and dotted lines have the same meanings as those in 
Fig.~\ref{te_nu}.
}
\label{te_nu2}
\end{figure}

Figure~\ref{te_nu} shows pulsation frequencies (cycle per day) of
excited modes versus effective temperature along the evolutionary
tracks of $15M_\odot$ and $20M_\odot$.  The left panel shows each
frequency and amplitude of the periodicities of HD 163899 detected by
the {\it MOST} satellite.  The horizontal dotted lines in the middle
and the right panels correspond to the {\it MOST} frequencies.  The
dashed lines show the frequency
given by $\sqrt{4\pi G\overline\rho}/(2\pi)$ along the evolutionary
track.  This line is used to distinguish between p-modes and g-modes
in Figure~\ref{hrd}.  Figure~\ref{te_nu2} shows similar plots for
$12M_\odot$, $13M_\odot$, and $14M_\odot$.

P-modes are excited in hotter models, while g-modes are excited in
relatively cooler ones.  The frequencies of the excited g-modes tend
to be higher for a larger latitudinal degree $l$.  In restricted
$T_{\rm eff}$ ranges, both p-modes and g-modes are excited
simultaneously, and the corresponding models are appropriate for HD
163899.  Since a linear stability analysis cannot predict the
amplitudes of unstable modes, we compare in this first attempt only
the predicted frequency ranges of excited modes with the observed ones.  
The observed frequency range is reproduced by the $20M_\odot$
models around $\log T_{\rm eff}\sim4.41$ and by the $15M_\odot$
models $\log T_{\rm eff} \sim 4.36$.  (These  are
indicated by vertical dash-dotted lines in Fig.~\ref{te_nu}.)  On the
other hand, for $14M_\odot$ and $13M_\odot$ models the frequency gap
between p-modes and g-modes looks too large to be consistent with the
observed frequency distribution.

Taking into account that the estimated range of $\log T_{\rm eff}$ for
B2Ib/II is roughly between 4.22 and 4.32 (Fig.~\ref{hrd}), a
$15M_\odot$ model at $\log T_{\rm eff} \approx 4.36$ looks best to
reproduce the observed frequency range of HD 163899.  Although the
frequency gap is still large, it would be filled, at least partially,
by rotational splittings.
(If a higher $T_{\rm eff}$ is allowed, $20M_\odot$ models at
$\log T_{\rm eff} \approx 4.41$ work better.)

In our models, rotation is completely disregarded.  In the presence of 
rotation a pulsation frequency measured in the observer's frame,
$\nu$(obs), shifts as
\begin{equation}
\nu({\rm obs}) = \nu({\rm rot}) - m\Omega
\end{equation}
from the frequency in the co-rotating frame, $\nu$(rot), where
$m$ is the azimuthal order ($|m|\le l$) and $\Omega$ is the rotation frequency.

If we assume that HD 163899 rotates at a speed of about 100km/s, an
average for B2Ib/II \citep{Abt02}, the rotation frequency would be
$\approx 0.1$c~d$^{-1}$.  If we assume $l \le 3$ for observed
pulsation modes, we expect that observed frequencies would be shifted
by at most $\approx \pm 0.3$ c~d$^{-1}$ from $\nu$(rot)s which we
assume to be close to those obtained for $\Omega=0$.  This effect
would nearly fill the frequency gap between p- and g-modes, and make
the predicted frequency distribution roughly consistent with the
observed one for HD 163899.

It is not possible to perform a detailed comparison between  
the detected frequencies and theoretical predictions until
accurate information on the parameters of HD 163899,
such as the effective temperature, 
the luminosity, and the rotation speed is available. 
The most important finding in our modeling exercises is the fact that
g-modes can be excited in post-main-sequence models, 
which suggests the presence of a new group of variable stars. 
We discuss, in the next section, the properties of excited g-modes. 

Figures~\ref{hrd} and \ref{te_nu} indicate that earlier B-supergiants 
should pulsate in p-modes. This theoretical
prediction is consistent with the recent observational result of
\citet{Kau06} who detected short-period line-profile variations
in HD 64760 (B0.5 Ib).   

Before we proceed to the next section, we note that there are
strongly-excited very-low frequency modes, which are most clearly seen
as a lowest-frequency branch in the frequency-$T_{\rm eff}$ diagram
for $20M_\odot$ in Figure~\ref{te_nu}.  The amplitude of these modes is
strongly confined to the exterior of the bottom of the Fe-convection
zone.  The frequencies are so low that these modes are propagative
even at the outer boundary.  Since we use a reflective boundary
condition at the outer boundary, we do not regard the stability of
these modes to be accurate.  Also, the property of these modes would
be modified significantly in the presence of slow rotation.
Therefore, we have not included these modes in Figure \ref{hrd}.  We
leave these modes for future investigations.

\section{EXCITATION OF G-MODES}
We have found that g-modes are excited in massive post-main-sequence
stars, whose central part is in radiative equilibrium.  This looks
contrary to the general thought that g-modes should be damped due to
very strong dissipation expected in a radiative core where the
Brunt-V\"ais\"al\"a frequency is very high.  The excitation becomes
possible, however, due to the presence of a fully-developed convective
shell associated with the hydrogen-burning shell in a massive star.
(Fully-developed convection means that convective mixing is strong
enough to homogenize the chemical composition.)  Since a g-mode
pulsation is evanescent in the convection zone, it can be reflected at
the boundary.  In other words, the convective shell can prevent a
g-mode from penetrating into the radiative core.

\begin{figure}
\epsscale{1.0}
\plotone{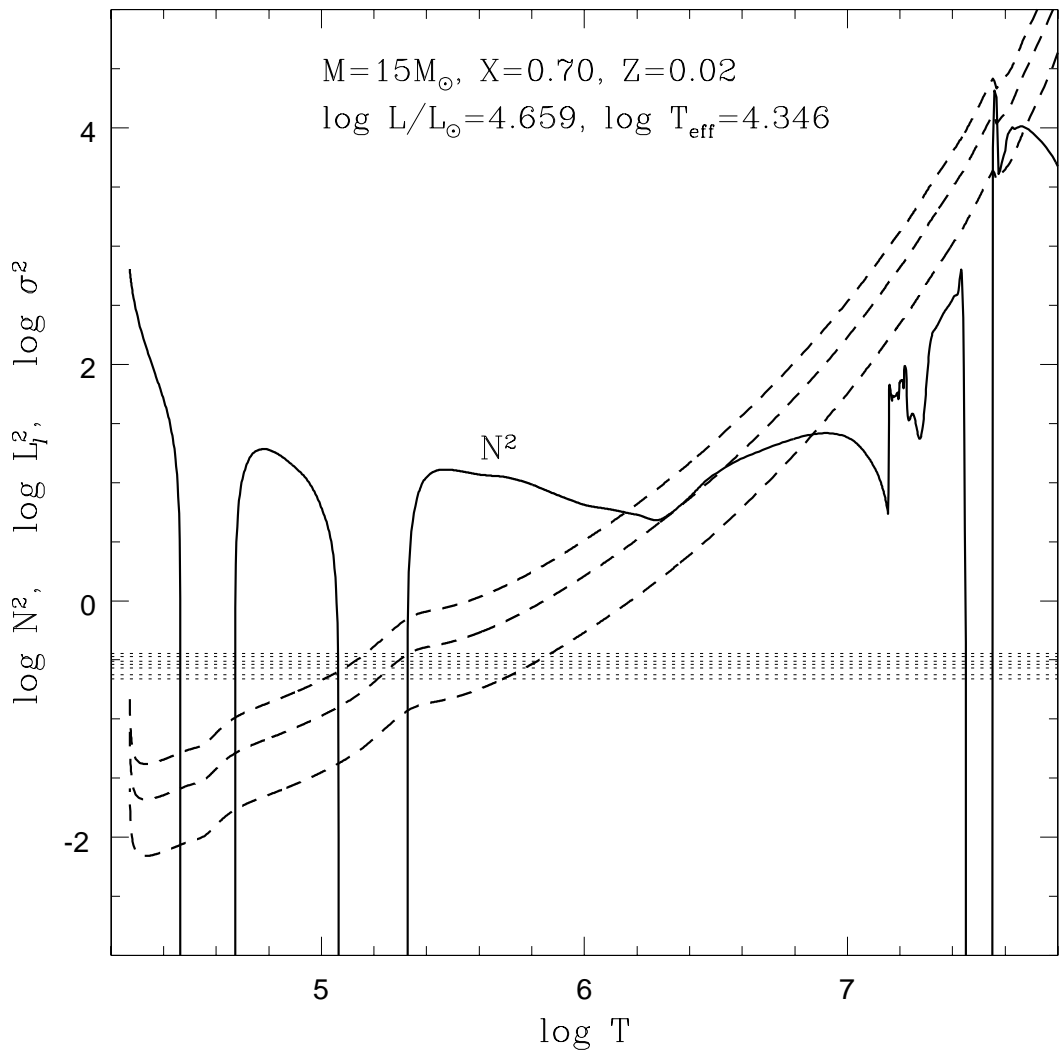}
\caption{
Runs of the square of the Brunt-V\"ais\"al\"a frequency, $N^2$, 
and the square of Lamb frequency $L^2_l=l(l+1)c_s^2/r^2$ for $l=1,2,3$.
Horizontal dotted lines indicate the angular frequencies ($\sigma$) 
of g-modes of $l=2$ excited in this model.
All the quantities in this figure are normalized by $GMR^{-3}$.
}
\label{prop}
\end{figure}

Figure~\ref{prop} shows runs of Lamb frequency 
$L_l^2 = l(l+1)c_s^2r^{-2}$, and Brunt-V\"ais\"ala 
frequency $N^2$, which is approximately written for an ideal gas as
\begin{equation}
N^2 \approx {g\over H_p}\left[\left(d\ln T\over d\ln P\right)_{\rm ad} + 
\left(d\ln\mu\over d\ln P\right) - \left(d\ln T\over d\ln P\right)
\right],
\label{eq_n2}
\end{equation}
where $c_s$ is the adiabatic sound speed, $r$ is the distance from the stellar
center, $g$ is the local gravity, $H_p$ is the pressure scale height,
$P$ is the pressure, and $\mu$ is the mean molecular weight.
Frequencies of excited g-modes of $l=2$ are shown by dotted horizontal lines.
In this figure all quantities are normalized by $GMR^{-3}$.

Roughly speaking, the radial wavenumber of a g-mode which pulsates
with an angular frequency $\sigma$ is proportional to
$\sigma^{-1}\sqrt{(N^2-\sigma^2)(L_l^2-\sigma^2)}$.  Therefore,
g-modes are radially propagative only if $\sigma^2 < N^2$ and
$\sigma^2 < L_l^2$, and evanescent otherwise \citep[see e.g.][for
details]{Unn89,Cox80}.  Since both $N^2$ and $L_l^2$ are very large in
the radiative core, we expect a very large wavenumber (i.e. very short
wavelength) and hence a large radiative damping for a g-mode in the
central part.  This is the reason why we do not generally expect
g-modes to be excited in a post-main-sequence star with a radiative
non-degenerate core.

In a massive post-main-sequence star, however, a fully developed
convection zone appears at the hydrogen-burning shell, which is seen
as a narrow gap around $\log T \sim 7.5$ in Figure~\ref{prop}.  This
zone, where g-modes are evanescent, can reflect some g-modes and
prevent them from penetrating into the core.  This effect reduces
radiative damping in the core significantly, and hence helps the
$\kappa$-mechanism in the envelope excite the mode.

\begin{figure}
\epsscale{1.0}
\plotone{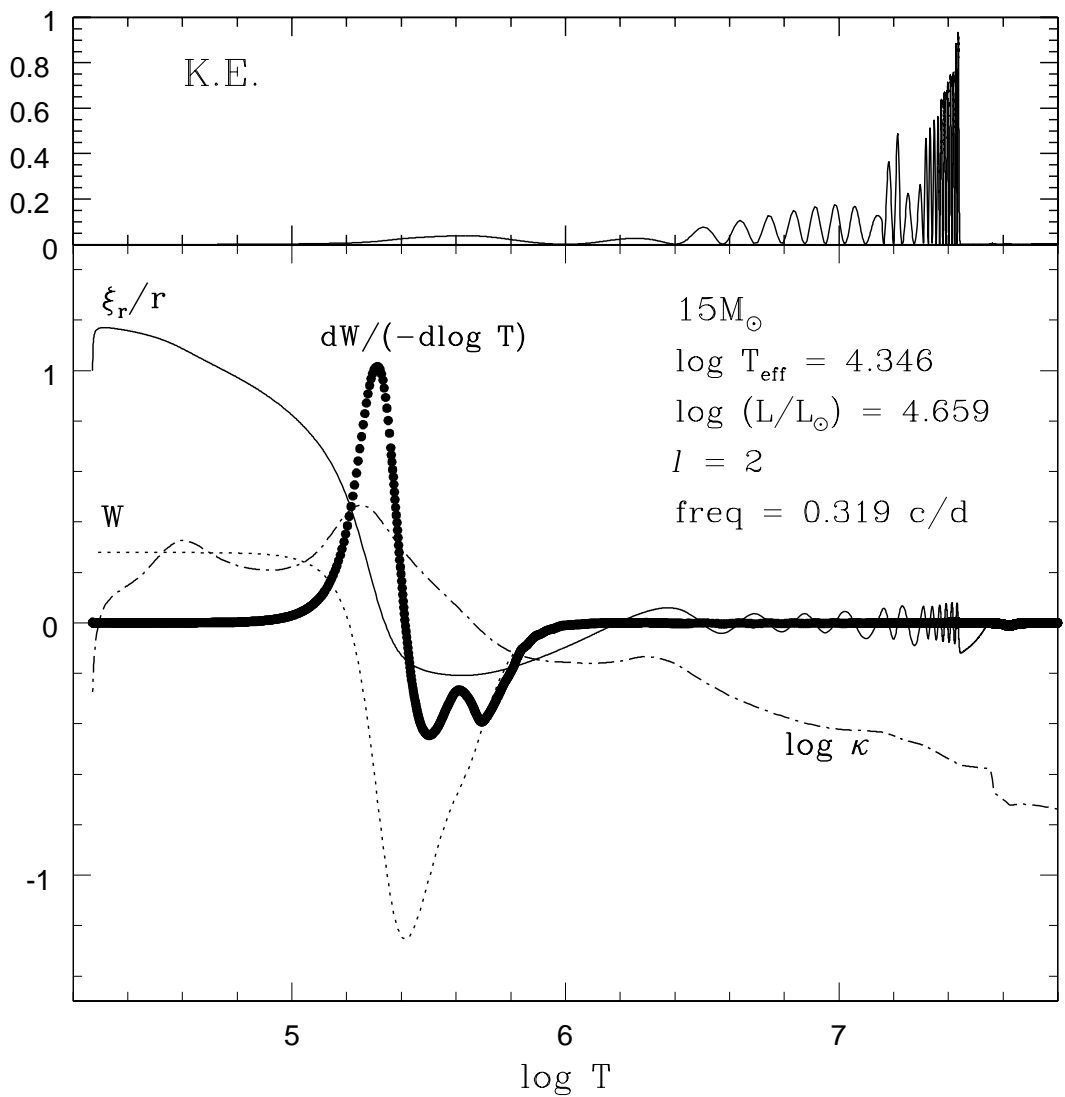}
\caption{
The bottom panel shows radial displacement (solid line), 
work $W$ (dotted line), and differential work $dW/(-d\log T)$ (dots), 
and opacity $\kappa$ (dash-dotted line) as a function of  $\log T$.
The top panel shows the distribution of the kinetic energy of the pulsation.
Radiative dissipation in the core is quenched because the pulsation
is reflected at the convective shell at 
$\log T\approx 7.5$. 
The opacity peak at $\log T \approx 5.3$ arises from Fe and that at 
$\log T \approx 4.6$ from HeII ionization.
}
\label{work}
\end{figure}

Figure~\ref{work} shows the work $W$, differential work $dW/d(-\log
T)$, radial displacement amplitude, and kinetic energy distribution
(top panel) of an excited g-mode.  The pulsation is excited if the
value of $W$ is positive at the surface.  ($dW/(-d\log T)>0$ means
that the zone drives pulsation.)  We see that the amplitude and the
kinetic energy suddenly becomes very small interior to the point at
$\log T \approx 7.5$, where the fully convective zone is located.
This indicates that the reflection at the convective shell actually
occurs for this mode.  Such a reflection at the convective shell is
essential for a g-mode to be excited in a post-main-sequence star by
the $\kappa$-mechanism at the Fe-bump of opacity.  All
post-main-sequence models which excite g-modes in Figure~\ref{hrd}
have a fully-developed convection zone associated with the
hydrogen-burning shell.  On the other hand, in a stable range of $4.16
\la \log T_{\rm eff} \la 4.20$ for $13M_\odot$ (Fig.~\ref{te_nu2}), for
example, no fully-developed convective zone is present at the
hydrogen-burning shell (i.e.  the whole convectively-unstable layers
are semiconvective only).

\begin{figure}
\epsscale{0.7}
\plotone{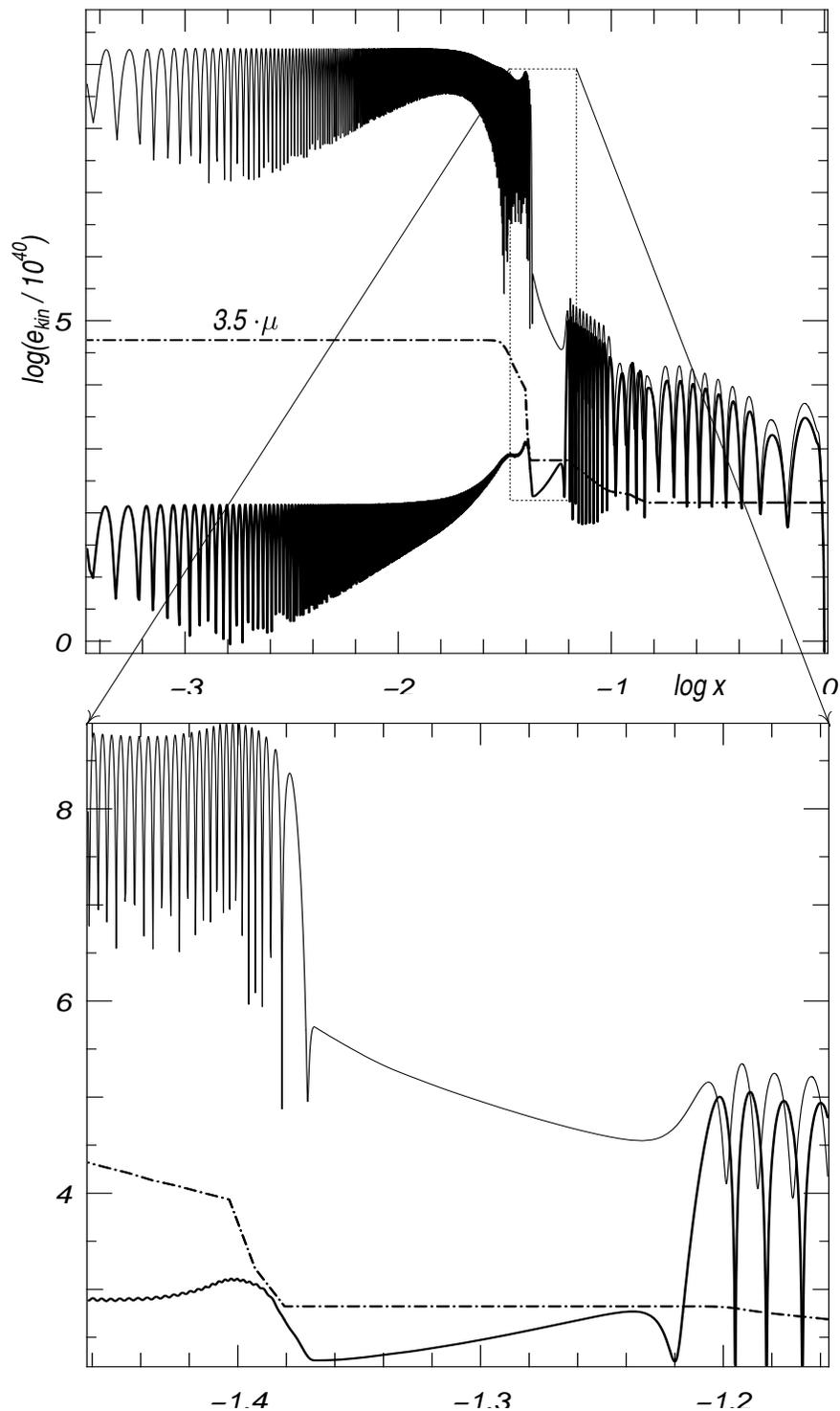}
\caption{
Kinetic energy distribution of an excited 0.3241 c~d$^{-1}$ mode (thick solid line)
and a damped 0.3256 c~d$^{-1}$ mode (thin solid line) of 
$l=2$ as a function of the fractional distance from the center ($x=r/R$). 
(The thin line is the upper one.)
Dot-dashed lines show the suitably scaled spatial variation 
of the mean molecular weight.
The kinetic energy of the excited mode is very small in the core
due to the reflection in the convective shell,
which reduces radiative damping in the core and hence makes the mode unstable.
}
\label{eigen}
\end{figure}

Only a few selected g-modes are reflected at the convection zone and
excited by the Fe-bump in an appropriate frequency range.
Figure~\ref{eigen} compares the kinetic energy distribution of an
excited $\ell=2$ g-mode (thick solid line) to that of a damped mode
(thin solid line) as a function of the fractional radius
($x=r/R$). (In this diagram the right hand side is the stellar
surface.)  The pulsation frequencies of the two modes are very close;
0.324 c~d$^{-1}$ and 0.326 c~d$^{-1}$.  These eigenfunctions were
calculated by the Riccati shooting method \citep{Gau96}, completely
resolving the spatial oscillations with roughly 16,500 integration
steps.  It is obvious that the kinetic energy in the core of the
excited mode is significantly reduced and hence the radiative damping
is quenched, while the kinetic energy of the damped mode is strongly
confined to the central part.

Most of the g-modes in the frequency range where excited modes reside
are confined to the radiative core and are damped due to strong
radiative damping.  Only those selected g-modes whose spatial
oscillations have the right phase at the boundary of the chemically
homogeneous shell convection zone are reflected and confined to the
envelope.  Therefore, which g-modes are excited depends on details of
the structure around the convective shell.  In the model shown in
Figure~\ref{prop} the layer in $7.23 \le \log T \le 7.44$ is
semi-convective (i.e. convective zone with an inhomogeneous chemical
composition).  We have treated semi-convective mixing based on the
model of \citet{Spr92}.  Everyone admits that many uncertainties lie
in the theory of convective mixing.  Since the Brunt-V\"ais\"al\"a
frequency depends sensitively on the distribution of chemical elements
[Eq.(\ref{eq_n2})], frequency spacings of the excited g-modes depend
on the treatment of convective mixing.  This indicates that comparing
observed frequencies with the predicted ones from models can constrain
the theory of convective mixing.  
We also note that the lower luminosity boundary of the g-mode 
instability region, which is the boundary of the presence of a fully 
developed shell convection, depends on the efficiency of convective mixing. 
Thus, g-modes in SPBsgs have great potential for asteroseismology.

As discussed above, the presence of a convective shell is crucial
to excite g-modes in the post-main-sequence models.  The appearance of
a convective shell in the post-main-sequence phase of massive stars
has been known for more than 30 years \citep[e.g.][]{Bar72}.  It has
also been known for a long time that g-modes can be reflected at the
boundary of a convection zone, and that the reflection quenches a
strong radiative dissipation in the core. Nevertheless, we can find no
published evidence of stability analyses of g-modes in massive
post-main-sequence stars since the emergence of OPAL opacities.  The
discovery of light variations in HD163899 by the {\it MOST} satellite,
which prompted us to perform stability analyses for such modes, played
a crucial role in our eventually finding the excited high-order g-modes.
On the HR diagram, the region with unstable g-modes seems to extend to
the region of $\alpha$ Cyg variables suggesting that g-mode pulsations
might play an important role in understanding these luminous
variable stars. The spatio-temporal evolution of the H-shell
convection zone in very massive stars during their post-main-sequence
evolution can be judged from figures in e.g. \citet{Lamb76} or
\citet{Maeder81}.

\section{CONCLUSION}
The {\it MOST} satellite discovered that the B type supergiant HD
163899 pulsates in g-modes as well as in p-modes, showing, for the
first time, SPB-type g-mode pulsations in a less luminous B supergiant.  
We have performed pulsation stability analyses for evolutionary models in the
mass ranges of $7\le M/M_\odot \le 20$, and found that g-modes are
excited by the Fe-bump of opacity in post-main-sequence stars in a
wide range of effective temperatures if the stellar mass is larger
than $\sim 12M_\odot$.  Our results indicate the existence of a
previously unrecognized group of variable stars; slowly pulsating B
supergiants (SPBsg) distinct from $\alpha$ Cyg variables.

The excitation of g-modes in B supergiants becomes possible by a
reflection of g-modes at the fully-developed convection zone
associated with the hydrogen-burning shell.  This reduces
significantly the radiative damping in the radiative core of the star.
Since the frequencies of excited g-modes depend on the structure
around the convective shell, SPBsg have a great potential for
asteroseismology.

HD 163899 is the first member of the SPBsg family.  Strictly speaking,
HD 163899, showing both g- and p-modes, is a hybrid of SPBsg and
$\beta$ Cephei-type.  The B supergiant $\iota$ CMa (B3 Ib/II) could be
another SPBsg member.  \citet{Bal85} obtained a 0.717 c~d$^{-1}$
periodicity for $\iota$ CMa.  Since the light curve looks irregular,
it is likely multi-periodic.  Another possible candidate could be HD
98410 (B2/B3 Ib/II) which is listed in \citet{Sta05} as one of the
``poor and rejected $\beta$ Cephei candidates'' with a
Hipparcos-deduced period of 1.453 days.  We also note that
\citet{Wae98}'s list for $\alpha$ Cyg-type supergiants includes some
less luminous B supergiants or bright giants such as HD54764
(B1Ib/II), HD141318 (B2II), and HD168183 (B1Ib/II).  These stars could
also be SPBsg candidates.  Further observations of these stars are
needed to confirm their membership.   Also, for a detailed
asteroseismological study of HD 163899, spectroscopic observations are
important to properly identify the pulsation modes that are
responsible for the photometrically observed frequencies.

\acknowledgments 
This research has made use of the SIMBAD
database, operated at CDS, Strasbourg, France.  The Natural Sciences
and Engineering Research Council of Canada supports the research of
D.B.G., J.M.M., A.F.J.M., S.M.R., E.S., G.A.H.W..  A.F.J.M. is also
supported by FQRNT (Qu\'ebec). R.K. is supported by the Canadian Space
Agency. W.W.W. is supported by the Austrian Space Agency and the
Austrian Science Fund (P14984).  H.S. is supported by the 21st Century
COE program of MEXT, Japan.



\clearpage

\begin{deluxetable}{rrrrr}
\tablewidth{310pt}
\tablecaption{HD 163899 periodicities from {\it MOST} photometry}

\tablehead{
  & \colhead{Freq[c~d$^{-1}$]} & \colhead{Amp[mmag]} & \colhead{phase[rad]\tablenotemark{a}} 
  & \colhead{S/N} }
\startdata
 f$_1$& 0.0431 $\pm$ 0.0003  &  4.03  $\pm$  0.08  &  1.77  $\pm$  0.04   & 19.2 \\
 f$_2$& 0.0726 $\pm$ 0.0003  &  3.00  $\pm$  0.06  &  3.97  $\pm$  0.04   & 11.6 \\
 f$_3$& 0.0211 $\pm$ 0.0009  &  0.91  $\pm$  0.09  &  5.12  $\pm$  0.12   & 10.2 \\
 f$_4$& 0.4711 $\pm$ 0.0003  &  2.32  $\pm$  0.03  &  4.34  $\pm$  0.03   & 11.9 \\
 f$_5$& 0.1911 $\pm$ 0.0002  &  2.37  $\pm$  0.03  &  6.05  $\pm$  0.02   & 10.9 \\
 f$_6$& 0.2246 $\pm$ 0.0003  &  1.96  $\pm$  0.03  &  1.48  $\pm$  0.03   & 11.0 \\
 f$_7$& 0.2840 $\pm$ 0.0002  &  1.52  $\pm$  0.06  &  6.63  $\pm$  0.02   &  8.5 \\
 f$_8$& 0.4009 $\pm$ 0.0003  &  1.69  $\pm$  0.03  &  1.16  $\pm$  0.04   &  8.4 \\
 f$_9$& 0.6931 $\pm$ 0.0009  &  0.91  $\pm$  0.04  &  6.68  $\pm$  0.12   &  7.5 \\
f$_{10}$& 0.5894 $\pm$ 0.0007  &  1.18  $\pm$  0.03  &  4.80  $\pm$  0.08   &  7.3 \\
f$_{11}$& 1.0789 $\pm$ 0.0003  &  1.24  $\pm$  0.03  &  4.90  $\pm$  0.05   &  7.1 \\
f$_{12}$& 0.1550 $\pm$ 0.0004  &  1.15  $\pm$  0.05  &  5.68  $\pm$  0.07   &  7.4 \\
f$_{13}$& 0.1035 $\pm$ 0.0006  &  1.10  $\pm$  0.03  &  3.14  $\pm$  0.08   &  6.3 \\
f$_{14}$& 0.3283 $\pm$ 0.0003  &  1.15  $\pm$  0.03  &  6.86  $\pm$  0.04   &  6.5 \\
f$_{15}$& 0.9683 $\pm$ 0.0011  &  0.72  $\pm$  0.03  &  5.25  $\pm$  0.13   &  6.0 \\
f$_{16}$& 0.6552 $\pm$ 0.0006  &  0.84  $\pm$  0.03  &  2.32  $\pm$  0.09   &  5.9 \\
f$_{17}$& 0.4279 $\pm$ 0.0006  &  1.03  $\pm$  0.03  &  1.64  $\pm$  0.07   &  6.0 \\
f$_{18}$& 1.9790 $\pm$ 0.0004  &  0.94  $\pm$  0.02  &  3.11  $\pm$  0.06   &  6.0 \\
f$_{19}$& 0.3626 $\pm$ 0.0006  &  0.94  $\pm$  0.03  &  4.06  $\pm$  0.07   &  5.6 \\
f$_{20}$& 1.3835 $\pm$ 0.0006  &  0.85  $\pm$  0.02  &  4.52  $\pm$  0.08   &  5.8 \\
f$_{21}$& 0.5010 $\pm$ 0.0008  &  0.81  $\pm$  0.03  &  3.25  $\pm$  0.11   &  5.4 \\
f$_{22}$& 1.1936 $\pm$ 0.0005  &  0.61  $\pm$  0.02  &  5.18  $\pm$  0.06   &  5.1 \\
f$_{23}$& 0.8182 $\pm$ 0.0005  &  0.95  $\pm$  0.03  &  4.60  $\pm$  0.07   &  5.0 \\
f$_{24}$& 0.2562 $\pm$ 0.0006  &  0.83  $\pm$  0.04  &  3.29  $\pm$  0.06   &  5.1 \\
f$_{25}$& 0.8798 $\pm$ 0.0009  &  0.97  $\pm$  0.06  &  6.62  $\pm$  0.10   &  5.0 \\
f$_{26}$& 0.7077 $\pm$ 0.0011  &  0.79  $\pm$  0.05  &  3.41  $\pm$  0.13   &  5.0 \\
f$_{27}$& 1.3391 $\pm$ 0.0010  &  0.69  $\pm$  0.03  &  6.38  $\pm$  0.13   &  4.7 \\
f$_{28}$& 1.4606 $\pm$ 0.0008  &  0.65  $\pm$  0.02  &  6.34  $\pm$  0.11   &  4.2 \\
f$_{29}$& 2.0495 $\pm$ 0.0004  &  0.56  $\pm$  0.02  &  5.80  $\pm$  0.07   &  4.5 \\
f$_{30}$& 0.8625 $\pm$ 0.0011  &  0.73  $\pm$  0.06  &  4.86  $\pm$  0.14   &  4.5 \\
f$_{31}$& 2.2975 $\pm$ 0.0010  &  0.46  $\pm$  0.02  &  2.29  $\pm$  0.12   &  4.1 \\
f$_{32}$& 0.9439 $\pm$ 0.0011  &  0.54  $\pm$  0.03  &  6.74  $\pm$  0.14   &  4.7 \\
f$_{33}$& 0.7648 $\pm$ 0.0009  &  0.59  $\pm$  0.03  &  5.37  $\pm$  0.11   &  4.4 \\
f$_{34}$& 1.6542 $\pm$ 0.0016  &  0.44  $\pm$  0.03  &  3.20  $\pm$  0.15   &  3.9 \\
f$_{35}$& 1.1444 $\pm$ 0.0007  &  0.50  $\pm$  0.02  &  6.39  $\pm$  0.08   &  4.1 \\
f$_{36}$& 1.9131 $\pm$ 0.0010  &  0.46  $\pm$  0.02  &  2.28  $\pm$  0.15   &  4.0 \\
f$_{37}$& 0.5628 $\pm$ 0.0016  &  0.41  $\pm$  0.04  &  4.53  $\pm$  0.19   &  3.9 \\
f$_{38}$& 2.5915 $\pm$ 0.0009  &  0.44  $\pm$  0.02  &  6.93  $\pm$  0.13   &  4.0 \\
f$_{39}$& 1.5739 $\pm$ 0.0008  &  0.53  $\pm$  0.03  &  6.96  $\pm$  0.10   &  4.0 \\
f$_{40}$& 0.1381 $\pm$ 0.0008  &  0.74  $\pm$  0.04  &  3.63  $\pm$  0.11   &  4.0 \\
f$_{41}$& 0.3047 $\pm$ 0.0004  &  0.58  $\pm$  0.06  &  5.46  $\pm$  0.05   &  3.9 \\
f$_{42}$& 1.3032 $\pm$ 0.0011  &  0.42  $\pm$  0.03  &  3.69  $\pm$  0.08   &  4.1 \\
f$_{43}$& 2.3823 $\pm$ 0.0007  &  0.44  $\pm$  0.02  &  5.93  $\pm$  0.11   &  4.0 \\
f$_{44}$& 2.8416 $\pm$ 0.0010  &  0.39  $\pm$  0.02  &  6.54  $\pm$  0.18   &  4.1 \\
f$_{45}$& 1.7444 $\pm$ 0.0014  &  0.39  $\pm$  0.02  &  3.31  $\pm$  0.12   &  3.9 \\
f$_{46}$& 1.5269 $\pm$ 0.0014  &  0.39  $\pm$  0.03  &  5.52  $\pm$  0.19   &  3.7 \\
f$_{47}$& 1.6172 $\pm$ 0.0014  &  0.46  $\pm$  0.03  &  7.17  $\pm$  0.18   &  3.6 \\
f$_{48}$& 1.0069 $\pm$ 0.0018  &  0.33  $\pm$  0.03  &  4.23  $\pm$  0.23   &  3.6 \\

\enddata
\tablenotetext{a}{
Phases are referenced to the first observation in the data set.
}
\end{deluxetable}

\end{document}